\documentclass[11pt]{article}
\usepackage{amssymb, latexsym, amsmath}
\usepackage{amsfonts, graphics}
\def\R{\mathbb R}

\def\N{\mathbb N}

\newcommand{\prfe}{\hspace*{\fill} $\Box$

\smallskip \noindent}

\begin{document}
\sloppy
\newtheorem{theorem}{Theorem}[section]
\newtheorem{definition}[theorem]{Definition}
\newtheorem{proposition}[theorem]{Proposition}
\newtheorem{example}[theorem]{Example}
\newtheorem{remark}[theorem]{Remark}
\newtheorem{cor}[theorem]{Corollary}
\newtheorem{lemma}[theorem]{Lemma}

\renewcommand{\theequation}{\arabic{section}.\arabic{equation}}

\title{Existence of axially symmetric static solutions
\\ of the Einstein-Vlasov system}

\author{H{\aa}kan Andr\'{e}asson\\
        Mathematical Sciences\\
        Chalmers University of Technology\\
        G\"{o}teborg University\\
        S-41296 G\"oteborg, Sweden\\
        email: hand@chalmers.se\\
        \ \\
        Markus Kunze\\
        Fakult\"at f\"ur Mathematik\\
        Universit\"at Duisburg-Essen\\
        D-45117 Essen, Germany\\
        email: markus.kunze@uni-due.de\\
        \ \\
        Gerhard Rein\\
        Fakult\"at f\"ur Mathematik, Physik und Informatik\\
        Universit\"at Bayreuth\\
        D-95440 Bayreuth, Germany\\
        email: gerhard.rein@uni-bayreuth.de}

\maketitle

\begin{abstract}
We prove the existence of static, asymptotically flat non-vacuum spacetimes
with axial symmetry where the matter is modeled as a collisionless gas.
The axially symmetric solutions of the resulting Einstein-Vlasov
system are obtained via the implicit function theorem by perturbing off
a suitable spherically symmetric steady state of the Vlasov-Poisson system.
\end{abstract}

\section{Introduction}
\setcounter{equation}{0}
The aim of the present investigation is to prove the existence of
static, asymptotically flat, and axially symmetric solutions of the 
Einstein-Vlasov system.
This system describes, in the context of general relativity,
the evolution of an ensemble of particles which
interact only via gravity. Examples from astrophysics of such
ensembles include galaxies or globular clusters where the stars
play the role of the particles and where collisions among these
particles are usually sufficiently rare to be neglected. 
The particle distribution is given by a density
function $f$ on the tangent bundle $TM$ of the spacetime manifold $M$.
We assume that all particles have the
same rest mass which is normalized to unity.
Hence the particle distribution 
function is supported
on the mass shell
\[
PM = \{ g_{\alpha \beta} p^\alpha p^\beta = -c^2\ 
\mbox{and}\ p^\alpha\ \mbox{is future pointing}\} 
\subset TM.
\]
Here $g_{\alpha \beta}$ denotes the Lorentz metric on the spacetime $M$
and if $x^\alpha$ are coordinates on $M$, then
$p^\alpha$ denote the corresponding canonical momentum coordinates; Greek
indices always run from $0$ to $3$, and we have a specific reason for 
making the dependence on the speed of light $c$ explicit.
We assume that the coordinates are chosen such that
\[
ds^2 = c^2 g_{00} dt^2 + g_{ab} dx^a dx^b
\] 
where Latin indices run from $1$ to $3$
and $t=x^0$ should be thought of as a timelike coordinate.
On the mass shell $p^0$ can be expressed by the remaining coordinates,
\[
p^0 = \sqrt{-g^{00}}\sqrt{1+ c^{-2} g_{ab}p^a p^b},
\]
and $f=f(t,x^a,p^b)\geq 0$. The Einstein-Vlasov system
now consists of the Einstein field equations
\begin{equation} \label{einst_gen}
G_{\alpha\beta} = 8 \pi c^{-4} T_{\alpha\beta}
\end{equation}
coupled to the Vlasov equation
\begin{equation} \label{vlasov_gen}
p^0 \partial_t f + p^a \partial_{x^a} f - 
\Gamma^a_{\beta \gamma} p^\beta p^\gamma \partial_{p^a} f = 0
\end{equation}
via the following definition of the energy momentum tensor:
\begin{equation} \label{emt_gen}
T_{\alpha \beta}
= c |g|^{1/2} \, \int p_\alpha p_\beta f \,\frac{dp^1 dp^2 dp^3}{-p_0}.
\end{equation}
Here $|g|$ denotes the modulus of the determinant of the metric, and 
$\Gamma^\alpha_{\beta \gamma}$ are the Christoffel symbols
induced by the metric.
We note that the characteristic system of the Vlasov equation
(\ref{vlasov_gen}) are the geodesic equations written as a first
order system on the mass shell $PM$ which is invariant under 
the geodesic flow.
For more background on the Einstein-Vlasov equation we refer to
\cite{And05}.

In \cite{Rein94,RR93,RR00} the existence of a broad variety
of static, asymptotically flat solutions of this system has been established,
all of which share the restriction that they are spherically
symmetric. The purpose of the present investigation is to
remove this restriction and prove the existence of 
static, asymptotically flat solutions to the Einstein-Vlasov system
which are axially symmetric but not spherically symmetric.
From the applications
point of view this symmetry assumption is more ``realistic'' 
than spherical symmetry, and from the mathematics point of view
the complexity of the Einstein field equations increases drastically
if one gives up spherical symmetry.

We use usual axial coordinates $t\in \R,\ \rho \in [0,\infty[,\
z\in \R,\ \varphi\in [0,2 \pi]$ and write 
the metric in the form
\begin{equation} \label{metric_ax}
ds^2=-c^2 e^{2\nu/c^2} dt^2 + e^{2\mu} d\rho^2 + e^{2\mu} dz^2+ 
\rho^2 B^2 e^{-2\nu/c^2} d\varphi^2 
\end{equation}
for functions $\nu, B, \mu$ depending on $\rho$ and $z$.
The reason for writing $\nu/c^2$ instead of $\nu$
is so that below $\nu$ converges to the Newtonian potential
$U_N$ in the limit $c\to \infty$. 
The metric is to be asymptotically flat in the sense
that the boundary values
\begin{equation} \label{bc_infinity}
\lim_{|(\rho,z)| \to \infty} \nu(\rho,z) = 
\lim_{|(\rho,z)| \to \infty} \mu(\rho,z) = 0,\
\lim_{|(\rho,z)| \to \infty} B(\rho,z) = 1
\end{equation}
are attained at spatial infinity with certain rates which are specified later.
In addition we need to require the condition that the metric is locally flat
at the axis of symmetry, i.e.,
\begin{equation} \label{bc_axis}
\nu(0,z)/c^2 + \mu(0,z) = \ln B(0,z),\ z\in \R .
\end{equation}
We refer to \cite{Bard} for more information on axially symmetric
spacetimes and state our main result.
\begin{theorem} \label{vaguemain}
There exist static solutions of the Einstein-Vlasov system
(\ref{einst_gen}), (\ref{vlasov_gen}), (\ref{emt_gen}) with $c=1$
such that 
the metric is of the form (\ref{metric_ax}) and satisfies the
boundary conditions (\ref{bc_infinity}), (\ref{bc_axis}),
and the spacetime is axially symmetric, but not spherically symmetric.
\end{theorem}
It should be pointed out that
the above form of the metric excludes solutions
with non-zero total angular momentum. Since the corresponding generalization
induces qualitatively new, additional difficulties it is postponed to
a later investigation.

The strategy of the proof of this result is as follows. 
Due to the symmetries of the metric the following quantities
are constant along geodesics:
\begin{eqnarray}
E 
&:=& - g(\partial/\partial t, p^\alpha) = c^2 e^{2\nu/c^2} p^0 \nonumber\\
&=& 
c^2 e^{\nu/c^2} \sqrt{1+c^{-2}\left(e^{2\mu} (p^1)^2 + e^{2\mu} (p^2)^2 + 
\rho^2 B^2 e^{-2\nu/c^2} (p^3)^2 \right)},\quad \label{Edef}\\
L 
&:=& 
g(\partial/\partial \varphi, p^\alpha) = 
\rho^2\,B^2 e^{-2\nu/c^2} p^3 \label{Ldef};
\end{eqnarray}
$E$ can be thought of as a local or particle energy and $L$ 
is the angular momentum of a particle with respect to the axis
of symmetry.
Since up to regularity issues a distribution function $f$ 
satisfies the Vlasov equation
if and only if it is constant along geodesics,
any distribution function $f$ which depends only on $E$ and $L$
satisfies the Vlasov equation with a metric of the above form. 
Hence we make the ansatz
\begin{equation} \label{ansatz_gen}
f(x^a,p^b) = \phi(E,L),
\end{equation}
and the Vlasov equation (\ref{vlasov_gen}) holds.
Upon insertion of this ansatz into the definition 
(\ref{emt_gen}) of the energy momentum tensor the latter
becomes a functional $T_{\alpha \beta} =  T_{\alpha \beta} (\nu,B,\mu)$
of the yet unknown metric functions
$\nu,B,\mu$, and we are left with the problem of solving the field
Einstein equations (\ref{einst_gen}) with this right hand side.
We obtain solutions by perturbing off spherically symmetric steady
states of the Vlasov-Poisson system via the implicit function theorem;
the latter system arises as the Newtonian limit of the Einstein-Vlasov
system. Our main result specifies conditions on the ansatz function
$\phi$ above such that a two parameter family of axially symmetric
solutions of the Einstein-Vlasov system passes through the corresponding
spherically symmetric, Newtonian steady state. The parameter
$\gamma = 1/c^2$ turns on general relativity and the second parameter $\lambda$
turns on the dependence on $L$ and hence axial symmetry;
notice that $L$ is not invariant under arbitrary rotations about
the origin, so if $f$ actually depends on $L$ the solution is
not spherically symmetric.
The scaling symmetry of the Einstein-Vlasov system can then 
be used to obtain the desired solutions 
for the physically correct value of $c$.

The detailed formulation of our result is stated in the next section
together with the basic set up of its proof. The remaining sections
of the paper are then devoted to establishing the various features of
the basic set up which are needed to apply the implicit function theorem,
and to prove various properties of the solutions we obtain.

We conclude this introduction with some further references to the
literature. The idea of using the implicit function theorem to
obtain equilibrium configurations of self-gravitating matter 
distributions from already known solutions can be traced back to
{\sc L.~Lichtenstein} who argued the existence of axially symmetric,
stationary, self-gravitating fluid balls in this way \cite{L1,L2}.
His arguments were put into a rigorous and modern framework 
in \cite{H}. The analogous approach was used in \cite{Rein00}
to obtain axially symmetric steady states of the Vlasov-Poisson system,
see also \cite{Schulze}.
The approach has also been used to construct axially symmetric stationary
solutions of the Einstein equations coupled to a matter model:
In \cite{H2} matter was described as an ideal fluid whereas
in \cite{ABS08,ABS09} is was described as a static
or a rotating elastic body respectively. 
Besides the different matter model our investigation differs
from the latter two in that we employ the rather explicit
form of the metric stated above and a reduced version
of the Einstein field equations which closely follows \cite{Bard}.


\section{Set up of the proof}
\setcounter{equation}{0}
\label{sec_res}


In what follows we also use the Cartesian coordinates
\[
(x^1,x^2,x^3)= (\rho\cos\varphi,\rho\sin\varphi, z) \in \R^3
\]
which correspond to the axial coordinates $\rho\in [0, \infty[$, $z\in\R$,
$\varphi \in [0,2 \pi]$; it should be noted that
tensor indices always refer to the spacetime coordinates $t,\rho,z,\varphi$.
By abuse of notation we write $\nu (\rho,z)=\nu(x)$ etc. 
In Section~\ref{regularity} we collect the relevant information
on the relation between regularity properties of axially symmetric functions
expressed in the variables
$x\in \R^3$ or $\rho\in [0, \infty[,\ z\in\R$, respectively.  

We introduce two (small) parameters $\gamma = 1/c^2 \in [0,\infty[$ 
and $\lambda\in \R$.
In order to obtain the correct Newtonian limit below we adjust
the ansatz for $f$ as follows. Let
\[
v^1 = e^{\mu} p^1,\ v^2 = e^{\mu} p^2,\ v^3 = \rho\,B e^{-\gamma \nu} p^3,
\]
so that
\[
p^0 = e^{-\gamma \nu} \sqrt{1+\gamma |v|^2}.
\]
For the particle distribution function we make the ansatz
\begin{equation} \label{fansatz}
f(x,v) 
= \phi\left(E-1/\gamma\right) \psi(\lambda L).
\end{equation}
The important point here is that
\begin{equation} \label{limitE}
E-1/\gamma = \frac{e^{\gamma \nu(x)}\sqrt{1+\gamma |v|^2} -1}{\gamma} 
\to \frac{1}{2} |v|^2 + \nu(x)
\ \mbox{as}\ \gamma \to 0, 
\end{equation}
i.e., the limit is the non-relativistic energy of a particle with phase space
coordinates $(x,v)$ in case $\nu=U_N$ is the Newtonian gravitational
potential. For $\gamma=0$ this limit is to replace the argument
of $\phi$ in (\ref{fansatz}).
We now specify the conditions on the functions $\phi$ and $\psi$.

\smallskip

\noindent
{\bf Conditions on $\phi$ and $\psi$.}
\begin{itemize}
\item[($\phi 1$)]
$\phi \in C^2(\R)$ and there exists $E_0>0$ such that
$\phi(\eta)=0$ for $\eta \geq E_0$ and $\phi(\eta) >0$ for $\eta < E_0$.
\item[($\phi 2$)]
The ansatz $f(x,v)=\phi\left(\frac{1}{2}|v|^2 + U(x)\right)$
leads to a compactly supported steady state of the Vlasov-Poisson 
system, i.e., there exists a solution $U=U_N \in C^2(\R^3)$
of the semilinear Poisson equation
\[
\Delta U = 4 \pi \rho_N =  
4 \pi \int\phi\left(\frac{1}{2}|v|^2 + U\right)\, dv,\
U(0)=0,
\]
$U_N(x) = U_N(|x|)$ is spherically symmetric, and the support of 
$\rho_N \in C^2(\R^3)$ is the closed ball
$\overline{B}_{R_N}(0)$ where $U_N(R_N)=E_0$ and  $U_N(r)<E_0$ 
for $0 \leq r < R_N$,
$U_N(r)>E_0$ for $r > R_N$.
\item[($\phi 3$)]
\hspace{2.8cm} $6+4\pi r^2 a_N(r)>0,\quad r\in [0, \infty[$,\\
where
\[
a_N(r):=\int_{\R^3}\phi'\Big(\frac{1}{2}\,|v|^2+U_N(r)\Big)\,dv.
\]
\item[($\psi$)]  
$\psi \in C^\infty(\R)$ is even with $\psi(L) = 1$ iff $L=0$,
and $\psi \geq 0$.
\end{itemize}
For such a steady state
\[
\lim_{|x|\to \infty} U_N(x) =U_N(\infty) > E_0.
\]
The normalization condition $U_N(0)=0$ instead of 
$U_N(\infty) =0$ is unconventional from the physics point
of view, but it has technical advantages below.
Examples for ansatz functions $\phi$ which satisfy
($\phi 1$) and ($\phi 2$) are found in \cite{BFH,RR00}, 
the most well-known ones being the polytropes
\begin{equation} \label{poly}
\phi(E):=(E_0-E)^k_+
\end{equation}
for $2<k<7/2$; here $E_0>0$ and $(\cdot)_+$ denotes the positive part.
In Section~\ref{an_cond} we show that for this class
of ansatz functions also ($\phi 3$) holds. Numerical
checks indicate that ($\phi 3$) holds for general isotropic
steady states of the Vlasov-Poisson system.

\smallskip

\noindent
We can now give a more detailed formulation of our result.
\begin{theorem} \label{main}
There exists $\delta >0$ and a two parameter
family 
\[
(\nu_{\gamma,\lambda},B_{\gamma,\lambda},\mu_{\gamma,\lambda})
_{(\gamma, \lambda) \in [0,\delta[\times ]-\delta,\delta [ } \subset C^2(\R^3)^3
\]
with the following properties:
\begin{itemize}
\item[(i)]
$(\nu_{0,0},B_{0,0},\mu_{0,0}) = (U_N,1,0)$ where $U_N$ is the potential
of the Newtonian steady state specified in $(\phi 2)$.
\item[(ii)]
If for $\gamma > 0$ a distribution function is defined
by Eqn.~(\ref{fansatz}) and a Lorentz metric by (\ref{metric_ax})
with $c=1/\sqrt{\gamma}$ then this defines a solution of the
Einstein-Vlasov system (\ref{einst_gen}), (\ref{vlasov_gen}), (\ref{emt_gen})
which satisfies the boundary condition (\ref{bc_axis}) and is 
asymptotically flat.
For $\lambda \neq 0$ this solution is not spherically symmetric.
\item[(iii)]
If for $\gamma=0$ a distribution function is defined
by Eqn.~(\ref{fansatz}), observing (\ref{limitE}), this yields
a steady state of the Vlasov-Poisson system with gravitational
potential $\nu_{0,\lambda}$ which is not spherically symmetric
for $\lambda \neq 0$.
\item[(iv)]
In all cases the matter distribution is compactly supported
both in phase space and in space.
\end{itemize}
\end{theorem}
{\bf Remark.}
\begin{itemize}
\item[(a)]
The smallness restriction to $\gamma=1/c^2$ is undesired because
$c$ is, in a given set of units, a definite number.
However, if $(f,\nu,B,\mu)$ is a static solution for some choice
of $c\in]0,\infty[$ then the rescaling
\begin{eqnarray*}
\tilde f(\rho,z,p^1,p^2,p^3)
&=&
c^{-3}
f(c \rho, c z,c p^1,c p^2,p^3),\\
\tilde \nu(\rho,z)
&=& 
c^{-2} \nu (c \rho, c z),\\
\tilde B(\rho,z)
&=& 
B (c \rho, c z), \\
\tilde \mu(\rho,z)
&=& 
\mu (c \rho, c z)
\end{eqnarray*}
yields a solution of the Einstein-Vlasov system with $c=1$.
The factor $c^2$ in the metric (\ref{metric_ax}) is removed by
a rescaling of time.
\item[(b)]
The smallness restriction to $\lambda$ means that the solutions
obtained are close to being spherically symmetric.
\item[(c)]
The metric does not satisfy the boundary conditions (\ref{bc_infinity}), but
\begin{equation} \label{bc_infinity_shift}
\lim_{|(\rho,z)| \to \infty} \nu(\rho,z) = \nu_\infty,\
\lim_{|(\rho,z)| \to \infty} \mu(\rho,z) = -\nu_\infty/c^2,\
\lim_{|(\rho,z)| \to \infty} B(\rho,z) = 1.
\end{equation}
However, if we by abuse of notation redefine $\nu=\nu-\nu_\infty$
and $\mu=\mu+\nu_\infty/c^2$ then the original
condition (\ref{bc_infinity}) is restored and
the metric (\ref{metric_ax}) takes the form
\begin{equation} \label{metric_ax_shift}
ds^2= -c^2 e^{2\nu/c^2} c_1^2dt^2 + 
c_2^2\left(e^{2\mu} d\rho^2 + e^{2\mu} dz^2+ 
\rho^2 B^2 e^{-2\nu/c^2} d\varphi^2 \right)
\end{equation}
with constants $c_1, c_2>0$ which simply amounts to
a choice of different units of time and space.
By general covariance of the Einstein-Vlasov system
(\ref{einst_gen}), (\ref{vlasov_gen}), (\ref{emt_gen})
the equations still hold. 
\item[(d)]
In view of \cite{Rein00} part (iii) of the theorem does not give
new information on steady states of the Vlasov-Poisson system
and is stated mainly in order to understand the obtained
two parameter family of states as a whole. However, we note
that for the Newtonian set-up in \cite{Rein00}
axially symmetric steady states were obtained as deformations
of a spherically symmetric one. The present approach
differs considerably from this and in principle
is more direct. 
\item[(e)]
In the course of the proof of the theorem additional regularity
properties and specific rates at which the boundary values
at infinity are approached will emerge.
\end{itemize}

In the rest of this section we transform the problem of finding
the desired solutions into the problem of finding zeros of
a suitably defined operator. The Newtonian steady state
specified in ($\phi 2$) will be a zero of this
operator for $\gamma=\lambda=0$, and the implicit function theorem
will yield our result. In order that the overall course
of the argument becomes clear we will go through its various steps,
postponing the corresponding detailed proofs to later sections.

The Einstein field equations are overdetermined, and we need to 
identify a suitable subset of (combinations of) these equations 
which, on the one hand, suffice to determine  $\nu, B, \mu$,
and which are such that at the end of the day all the field equations
hold once this reduced system is solved. 
We introduce the auxiliary metric function 
\[
\xi = \gamma \nu + \mu.
\]
Let $\Delta$ and $\nabla$ denote the Cartesian Laplace and gradient operator
respectively. Taking suitable combinations of the field equations one finds
that 
\begin{eqnarray}
&&
\Delta \nu + \frac{\nabla B}{B} \cdot \nabla \nu 
=
4\pi \gamma\left[\gamma e^{(2\xi - 4\gamma \nu)} T_{00} + T_{11} + T_{22}
+ \frac{1}{\rho^2 B^2} e^{2\xi}T_{33}\right],
\qquad \label{nu_eqn}\\
&&
\Delta B + \frac{\nabla \rho}{\rho}\cdot \nabla B = 
8 \pi \gamma^2 B \left( T_{11} + T_{22}\right) ,\label{B_eqn}\\
&&
\left(1+\rho \frac{\partial_\rho B}{B}\right)\partial_\rho \xi - 
\rho \frac{\partial_z B}{B} \partial_z \xi \nonumber \\
&&
\qquad\qquad
= \frac{1}{2\rho B} \partial_\rho(\rho^2 \partial_\rho B) - 
\frac{\rho}{2 B}\partial_{zz} B + \gamma^2 \rho\left((\partial_\rho \nu)^2 - 
(\partial_z \nu)^2\right),
\qquad \label{xi_eqna}\\
&&
\left(1+\rho \frac{\partial_\rho B}{B}\right)\partial_z \xi + 
\rho\frac{\partial_z B}{B} \partial_\rho \xi
= \frac{\partial_\rho(\rho \partial_z B)}{B} + 
2 \gamma^2 \rho\, \partial_\rho \nu \partial_z \nu .\label{xi_eqnb}
\end{eqnarray}
The last two equations arise from $\rho\, (G_{11} -G_{22})=0$
and $\rho\, G_{12} =0$ respectively;
note that due to (\ref{fansatz}), $T_{11} = T_{22}$ and $T_{12}=0$.
Because of the asymptotic behavior of $B$ and the structure of the
left hand side of (\ref{B_eqn}) we write
\[
B = 1 + h/\rho.
\]
Next, we observe that by taking suitable combinations of
(\ref{xi_eqna}) and (\ref{xi_eqnb}) we obtain equations
which contain only $\partial_\rho \xi$ or $\partial_z \xi$ 
respectively, and we chose the former.
In the above equations
the terms $T_{\alpha \beta}$ are functions of the unknown quantities
$\nu, h, \xi=\gamma \nu + \mu$ for which we therefore
have obtained the following reduced system of equations:
\begin{equation}
\Delta \nu = 
4\pi \left(\Phi_{00}
+ \gamma \Phi_{11} + \gamma \Phi_{33}\right)( \nu,B,\xi,\rho;\gamma,\lambda)
- \frac{1}{B} \nabla (h/\rho) \cdot \nabla \nu,
\quad \label{rnu_eqn}
\end{equation}
\begin{equation}
\partial_{\rho\rho} h + \partial_{zz} h=
8\pi \gamma^2 \rho B \Phi_{11} ( \nu,B,\xi,\rho;\gamma,\lambda),
\label{rh_eqn}
\end{equation}
\begin{eqnarray}
&&
\left((1+\partial_\rho h)^2 + (\partial_z h)^2\right)\, \partial_\rho \xi
=
\partial_z h \left(\partial_{z\rho}h + 
2 \gamma^2 (\rho+h)\partial_\rho \nu \partial_z \nu\right)\nonumber\\
&&
\qquad
{}+(1+\partial_\rho h)
\left(\frac{1}{2} (\partial_{\rho\rho}h - \partial_{zz} h) 
+\gamma^2(\rho + h)
\left((\partial_\rho \nu)^2 - (\partial_z \nu)^2\right)\right).\qquad\
\label{rxi_eqn}
\end{eqnarray}
We supplement this with the boundary condition (\ref{bc_axis}) 
which in terms of the 
new unknowns and since necessarily $h(0,z)=0$, reads
\begin{equation} \label{bcxi_axis}
\xi(0,z)=\ln\left(1+\partial_\rho h(0,z)\right).
\end{equation}
It remains to determine precisely the dependence of the
functions $\Phi_{\alpha\beta}$ on the unknown quantities
$\nu, h, \xi$.
Since the ansatz (\ref{fansatz}) is even in the momentum
variables $p^1, p^2, p^3$---the fact that $\psi$ is even is 
needed here---, all the off-diagonal elements of the energy-momentum
tensor vanish.
The computation of its non-trivial components
uses the new integration variables
\[
\eta = \frac{e^{\gamma \nu}\sqrt{1+\gamma |v|^2} -1}{\gamma},\ s = v^3,
\]
the abbreviation
\[
m(\eta,B,\nu,\gamma) = 
B e^{-\gamma \nu}\sqrt{\frac{e^{-2\gamma \nu}(1+\gamma \eta)^2 -1}{\gamma}},
\]
and yields
\begin{eqnarray}
&&
\Phi_{00}( \nu,B,\xi,\rho;\gamma,\lambda)
=
\gamma^2 e^{(2\xi - 4\gamma \nu)} T_{00}\label{Phi00}\\
&&
\qquad =
\frac{4\pi}{B} e^{(2\xi - 4\gamma \nu)}
\int_{(e^{\gamma \nu}-1)/\gamma}^\infty \phi(\eta) (1+\gamma \eta)^2 
\int_0^{m(\eta,B,\nu,\gamma)}\psi(\lambda \rho s)\, ds\, d\eta,\nonumber\\
&&
\Phi_{11}( \nu,B,\xi,\rho;\gamma,\lambda)
=
T_{11}+T_{22}\label{Phi11}\\
&&
\qquad =
\frac{4\pi}{B^3}e^{2\xi}
\int_{(e^{\gamma \nu}-1)/\gamma}^\infty \phi(\eta)
\int_0^{m(\eta,B,\nu,\gamma)}\psi(\lambda \rho s)
(m^2(\eta,B,\nu,\gamma)-s^2)\, ds\, d\eta,\nonumber\\
&&
\Phi_{33}( \nu,B,\xi,\rho;\gamma,\lambda)
=
\frac{e^{2\xi}}{\rho^2 B^2} T_{33}\label{Phi33}\\
&&
\qquad =
\frac{4\pi}{B^3}e^{2\xi}
\int_{(e^{\gamma \nu}-1)/\gamma}^\infty \phi(\eta)
\int_0^{m(\eta,B,\nu,\gamma)}\psi(\lambda \rho s)\,s^2 ds\, d\eta;\nonumber
\end{eqnarray} 
we recall that $T_{11}=T_{22}$.
The reason for keeping $B$ as argument on the right hand sides
above is that the matter terms are differentiable in this variable,
but taking a derivative with respect to $h$ would yield an irritating 
factor $1/\rho$. For elements of the function space chosen below
$h/\rho$ extends smoothly to the axis of symmetry $\rho=0$. 

We now define the function spaces in which we will obtain
the solutions of the system (\ref{rnu_eqn}), (\ref{rh_eqn}), (\ref{rxi_eqn}).
As noted above we write, by abuse of notation, axially symmetric
functions as functions of $x\in \R^3$ or of $\rho\geq 0, z\in \R$;
regularity properties of axially symmetric functions are considered
in Section~\ref{regularity}.
We fix $0 < \alpha< 1/2$ and $0 < \beta<1$, and consider the Banach spaces
\begin{eqnarray*}
{\cal X}_1
&:=& 
\Big\{\nu \in C^{3, \alpha}(\R^3) \mid 
\nu(x)=\nu(\rho,z)=\nu(\rho,-z)\ \mbox{and}\ 
\|\nu\|_{{\cal X}_1} <\infty\Big\},\\
{\cal X}_2
&:=& 
\Big\{h\in C^{4, \alpha}(\R^2) \mid 
h(\rho,z)=-h(-\rho,z)=h(\rho,-z)\ \mbox{and}\ 
\|h\|_{{\cal X}_2} <\infty\Big\},\\
{\cal X}_3
&:=& 
\Big\{\xi\in C^{2, \alpha}(Z_R) \mid
\xi(x)=\xi(\rho,z)=\xi(\rho,-z)\ \mbox{and}\ 
\|\xi\|_{{\cal X}_3} <\infty\Big\},
\end{eqnarray*}
where 
\[
Z_R:= \{ x\in \R^3 \mid \rho < R\}
\]
is the cylinder of radius $R>0$, the latter being defined in (\ref{Rdef})
below.
The norms are defined by
\begin{eqnarray*}
\|\nu\|_{{\cal X}_1} 
&:=&
\|\nu\|_{C^{3, \alpha}(\R^3)} 
+ \|(1+|x|)^{1+\beta}\nabla \nu\|_\infty, \\ 
\|h\|_{{\cal X}_2}
&:=& 
\|h\|_{C^{4, \alpha}(\R^2)} + \|(1+|(\rho,z)|)^{2}\nabla (h/\rho)\|_\infty,\\
\|\xi\|_{{\cal X}_3}
&:=& 
\|\xi\|_{C^{2, \alpha}(Z_R)},
\end{eqnarray*}
and
\[
{\cal X} := {\cal X}_1\times {\cal X}_2\times {\cal X}_3,\qquad
\|(\nu, h, \xi)\|_{\cal X} := \|\nu\|_{{\cal X}_1}+\|h\|_{{\cal X}_2}
+ \|\xi\|_{{\cal X}_3}.
\]
Here $\|\cdot\|_\infty$ denotes the
$L^\infty$-norm, functions in $C^{k, \alpha}(\R^n)$
have by definition continuous derivatives up to order $k$
and all the highest order derivatives are H\"older continuous with exponent
$\alpha$,
\[
\|g\|_{C^{k, \alpha}(\R^n)} 
:= 
\sum_{|\sigma|\leq k} \|D^\sigma g\|_\infty
+
\sum_{|\sigma| =  k}
\sup_{x, y\in \R^n, x\neq y} \frac{|D^\sigma g(x)-D^\sigma g(y)|}{|x-y|^\alpha}.
\]
and $D^\sigma$ denotes the derivative
corresponding to a multi-index $\sigma \in \N_0^n$. 
We note that if $h\in {\cal X}_2$ then $B=1+h/\rho \in C^3(\R^3)$,
cf.\ Lemma~\ref{hoverrhoreg}. Moreover, it will
be straightforward to extend $\xi$ to $\R^3$ once
a solution is obtained in the above space.

Now we recall the properties of the Newtonian
steady state specified in ($\phi 2$).
That condition implies that there exists $R>R_N>0$ such that 
\begin{equation}\label{Rdef}
U_N(r) > (E_0+U_N(\infty))/2,\ r > R.
\end{equation}
If 
\[
|| \nu - U_N ||_\infty < |E_0-U_N(\infty)|/4\ \mbox{and}\ 0\leq \gamma < \gamma_0,
\]
with $\gamma_0>0$ sufficiently small, depending on $E_0$ and $U_N$,
then
\[
\frac{e^{\gamma \nu(x)} -1}{\gamma} > E_0\ \mbox{for all}\ |x| > R.
\]
This implies that there exists some $\delta >0$ such that
for all $(\nu,h,\xi;\gamma, \lambda)\in {\cal U}$
the matter terms resulting from (\ref{Phi00})--(\ref{Phi33})
are compactly supported in $B_R(0)$, where
\[
{\cal U}:=\{(\nu,h,\xi;\gamma, \lambda) \in  {\cal X}\times [0,\delta[
\times ]-\delta,\delta[
\;\mid\; \|(\nu,h,\xi)-(U_N,0,0)\|_{\cal X} < \delta\}.
\]
In addition we require that $\delta>0$ is sufficiently small
so that for all elements
in ${\cal U}$ it holds that $B=1+h/\rho > 1/2$,
and the factor in front of $\partial_\rho \xi$ in (\ref{rxi_eqn})
is larger than $1/2$; since $h$ vanishes on the
axis of symmetry, $h/\rho$ is controlled by $\nabla h$. Now let an element
$(\nu,h,\xi;\gamma, \lambda) \in {\cal U}$ be given and substitute
it into the matter terms defined in (\ref{Phi00})--(\ref{Phi33}).
With the right hand sides obtained in this way the equations
(\ref{rnu_eqn})--(\ref{rxi_eqn}) can then be solved, 
observing the boundary condition
(\ref{bcxi_axis}) and the fact that we require $h$ 
to vanish on the axis of symmetry.
We define the corresponding solution operators by
\begin{eqnarray*}
G_1(\nu, h, \xi; \gamma, \lambda)(x)
&:=&
-\int_{\R^3} \left(\frac{1}{|x-y|}-\frac{1}{|y|}\right)\,M_1(y)\,dy\\
&&
{}+ \frac{1}{4 \pi}\int_{\R^3}
\frac{\nabla(h/\rho)(y)\cdot\nabla \nu(y)}{B(y)}\,\frac{dy}{|x-y|},\\
G_2(\nu, h, \xi; \gamma, \lambda)(x)
&:=&
4 \int_{\R^2}
\ln |(\rho-\tilde{\rho}, z-\tilde{z})|\,\tilde{\rho}
\,M_2(\tilde{\rho}, \tilde{z})\,d\tilde{\rho}\,d\tilde{z},\\
G_3(\nu, h, \xi; \gamma, \lambda)(x)
&:=&
\ln\left(1+ \partial_\rho h(0,z)\right)
+ \int_0^\rho g(s,z)\,ds,\ 0\leq \rho < R.
\end{eqnarray*}
Here
\begin{eqnarray*}
M_1(x) 
&:=& 
(\Phi_{00}+\gamma\Phi_{11}+\gamma\Phi_{33})
(\nu(x), B(x), \xi(x),\rho,; \gamma, \lambda),\\ 
M_2(\rho,z) 
&:=& 
\gamma^2 B(x)\,
\Phi_{11}(\nu(x), B(x), \xi(x),\rho; \gamma, \lambda),
\end{eqnarray*}
$M_2(\rho,z)=M_2(-\rho,z)$ for $\rho < 0$ and $z\in \R$,
and
\begin{eqnarray} \label{gdef}
g 
&:=& 
\left((1+\partial_\rho h)^2 + (\partial_z h)^2\right)^{-1}
\Bigg[\partial_z h \left(\partial_{z\rho}h + 
2 \gamma^2 (\rho+h)\partial_\rho \nu \partial_z \nu\right)\nonumber \\
&&
{}+
(1+\partial_\rho h)
\left(\frac{1}{2} (\partial_{\rho\rho}h - \partial_{zz} h) 
+\gamma^2(\rho + h)
\left((\partial_\rho \nu)^2 - (\partial_z \nu)^2\right)\right)\Bigg].\qquad\ 
\end{eqnarray}
Finally we define the mapping to which we are going to apply the
implicit function theorem as
\[ 
{\cal F}: {\cal U} \to {\cal X},\
(\nu, h, \xi; \gamma, \lambda)\mapsto
(\nu, h, \xi)-(G_1,G_2,G_3)(\nu, h, \xi; \gamma, \lambda).
\]
The proof of Theorem~\ref{main} now proceeds in a number of steps.

\noindent
{\em Step 1.}\\
As a first step we need to check
that the mapping ${\cal F}$ is well defined, in particular it preserves 
the various regularity and decay assumptions. This is done 
in Section~\ref{Fwelldef}.

\noindent
{\em Step 2.}\\
The next step is to see that 
\[
{\cal F}(U_N, 0, 0; 0, 0)=0. 
\]
This is due to the fact that for $\gamma=\lambda=0$ the choice $h=\xi=0$
trivially satisfies (\ref{rh_eqn}), (\ref{rxi_eqn}), while (\ref{rnu_eqn})
reduces to
\[
\Delta \nu =
4\pi \Phi_{00}(\nu,1,0;0,0)
\]
with
\[
\Phi_{00}(\nu,1,0;0,0) = 4\pi \int_{\nu}^\infty \phi(\eta)  
\sqrt{2(\eta - \nu)}\,d\eta = 
\int_{\R^3} \phi\left(\frac{1}{2}|v|^2 + \nu\right)\, dv;  
\]
notice that $h=0$ implies that $B=1$. By ($\phi 2$), $\nu = U_N$
is a solution of this equation, and the fact that $U_N\in {\cal X}_1$
is part of what was shown in the previous step.

\noindent
{\em Step 3.}\\
Next we show that ${\cal F}$ is continuous, and continuously Fr\'{e}chet
differentiable with respect to $(\nu, h, \xi)$. 
The fairly technical but straightforward details
are covered in Section~\ref{Fdiff}. 

\noindent
{\em Step 4.}\\
The crucial step is to see that the Fr\'{e}chet derivative
\[
L:= D{\cal F}(U_N, 0, 0; 0, 0): {\cal X} \to {\cal X}
\]
is one-to-one and onto. Indeed, 
\[
L(\delta\nu,\delta h,\delta\xi) 
= \left(\delta\nu-L_1(\delta\nu)-L_2(\delta h),\delta h,
\delta\xi -L_3(\delta h)\right)
\]
where
\begin{eqnarray*}
L_1(\delta\nu)(x)
&:=&
-\int_{\R^3} \left(\frac{1}{|x-y|}-\frac{1}{|y|}\right)\,
 a_N (y) \delta\nu (y)\,dy,\\
L_2(\delta h)(x)
&:=&
\frac{1}{4 \pi}\int_{\R^3}
\nabla(\delta h/\rho)(y)\cdot\nabla U_N(y)\,\frac{dy}{|x-y|},\\
L_3(\delta h)(x)
&:=& 
\partial_\rho\delta h (0,z) + 
\frac{1}{2}
\int_0^\rho (\partial_{\rho\rho} \delta h-\partial_{zz} \delta h)(s,z)\, ds,
\ 0\leq \rho < R,
\end{eqnarray*}
with $a_N$ as defined in ($\phi 3$).
To see that $L$ is one-to-one let
$L(\delta\nu,\delta h,\delta\xi)=0$.
Then the second component of this identity implies that
$\delta h =0$, and hence also $\delta\xi=0$ by the third component.
It therefore remains to show that
$\delta\nu=0$ is the only solution of the equation 
$\delta\nu = L_1(\delta\nu)$, i.e., of the equation
\begin{equation} \label{kernel} 
\Delta \delta\nu = 4 \pi a_N \delta\nu,\ \delta\nu(0)=0
\end{equation}
in the space ${\cal X}_1$. Under the assumption on $a_N$
stated in ($\phi 3$) this is correct and shown in 
Section~\ref{onetooneonto}.
It is at this point that
our unconventional normalization condition in $(\phi 2)$
together with the shift in the solution operator $G_1$
become important; notice that $L_1(\delta\nu)(0)=0$.

To see that $L$ is onto let $(g_1,g_2,g_3) \in {\cal X}$
be given. We need to show that there exists 
$(\delta\nu,\delta h,\delta\xi) \in {\cal X}$
such that $L(\delta\nu,\delta h,\delta\xi)=(g_1,g_2,g_3)$. The second
component of this equation simply says that $\delta h=g_2$. 
Now $\delta h \in {\cal X}_2$ implies that $L_3(\delta h) \in {\cal X}_3$, 
cf.~Lemma~\ref{axreg}~(b).
Hence we set $\delta \xi = g_3 +L_3(\delta h)$ to satisfy the third component
of the onto equation, and it remains to show that the equation
\begin{equation}\label{onto}
\delta\nu - L_1(\delta\nu) = g_1 + L_2(\delta h)
\end{equation}
has a solution $\delta\nu \in {\cal X}_1$.
Firstly, $L_2(\delta h)\in {\cal X}_1$. 
The assertion therefore follows from the fact that
$L_1 :{\cal X}_1 \to {\cal X}_1$ is compact, as is shown in 
Lemma~\ref{compactness}.  
 
We are now ready to apply the implicit function theorem,
cf.\ \cite[Thm.~15.1]{deimling}, to the mapping
${\cal F}: {\cal U} \to {\cal X}$; strictly speaking we should
suitably extend ${\cal F}$ to $\gamma < 0$, but this is not
essential. We obtain the following result.
\begin{theorem}\label{reducthm} 
There exists
$\delta_1, \delta_2 \in ]0, \delta[$ and a unique, continuous solution map
\[ 
S: [0, \delta_1[\times ]-\delta_1, \delta_1[
\to B_{\delta_2}(U_N,0,0)\subset {\cal X} 
\]
such that $S(0, 0)=(U_N, 0, 0)$ and
\[ 
{\cal F}(S(\gamma, \lambda); \gamma, \lambda)=0\ 
\mbox{for all}\ 
(\gamma, \lambda) \in [0, \delta_1[ \times ]-\delta_1, \delta_1[.
\]
\end{theorem}
The definition of ${\cal F}$ implies that for any 
$(\gamma,\lambda)$ the
functions $(\nu,h,\xi)=S(\gamma,\lambda)$ are a solution
of the equations (\ref{rnu_eqn})--(\ref{rxi_eqn}), and if
$f$ is defined by (\ref{fansatz}) then the equations
(\ref{nu_eqn}), (\ref{B_eqn}), (\ref{rxi_eqn}) hold with the induced
energy momentum tensor. We can extend $\xi$ to the whole space
using the solution operator $G_3$ for all $x\in \R^3$.
Also, the boundary condition
(\ref{bc_axis}) on the axis of symmetry is satisfied:
\[
\xi(0,z)= G_3(\nu,h,\xi)(0,z) = \ln (1+\partial_\rho h(0,z))= \ln B(0,z);
\] 
recall that $\xi = \gamma \nu + \mu$.
For $\gamma=0$ we conclude first that $h=0$, cf.~(\ref{rh_eqn})
or the $G_2$-part of the solution operator respectively,
then the $G_3$-part implies that $\xi=0$ so that the
solution reduces to $(\nu,0,0)$ where $\nu$
solves
\[
\Delta \nu = 4 \pi \Phi_{00}(\nu,1,0,\rho;0,\lambda).
\]
Since 
\[
\Phi_{00}(\nu,1,0,\rho;0,\lambda) = 4\pi
\int_{\nu}^\infty \phi(\eta)  
\int_0^{\sqrt{2(\eta - \nu)}} \psi(\lambda \rho s)\, ds\, d\eta
\]
coincides with the spatial density
induced by the ansatz (\ref{fansatz})
for the Newtonian case, cf.\ \cite[Lemma~2.1]{Rein00},
part (iii) of Theorem~\ref{main} is established.
If $\lambda\neq 0$ then condition ($\psi$) implies that $f$
really depends on the angular momentum variable $L$
which is not invariant under all rotations about the origin,
but only invariant under rotations about the axis $\rho=0$.
Moreover, if the metric were spherically symmetric then
the explicit dependence of the quantities
$\Phi_{jj}$ on $\rho$ would imply that the induced energy
momentum tensor would not be spherically symmetric
which is a contradiction. Hence the obtained solutions
are not spherically symmetric if  $\lambda\neq 0$. 
To complete the proof of Theorem~\ref{main}
we must show that indeed {\em all} the field equations
are satisfied by the obtained metric (\ref{metric_ax}).
The corresponding argument relies on the Bianchi identity
$\nabla_\alpha G^{\alpha \beta}=0$ which holds for the
Einstein tensor induced by any (sufficiently regular) metric,
and on the identity $\nabla_\alpha T^{\alpha \beta}=0$
which is a direct consequence of the Vlasov equation~(\ref{vlasov_gen});
$\nabla_\alpha$ denotes the covariant derivative corresponding 
to the metric (\ref{metric_ax}).
The details are carried out in Section~\ref{allfield}.

Finally we collect the additional information
on the solution which we obtain in the course of the proof.
\begin{proposition}\label{moreinfo}
Let $(\nu,h,\xi)=S(\gamma,\lambda)$ be any of the solutions
obtained in Theorem~\ref{reducthm} and define
$\mu := \xi -\nu/c^2$ and $B=1+h/\rho$. Then the limit
$\nu_\infty:=\lim_{|x|\to\infty} \nu(x)$ exists, and
for any $\sigma\in\N_0^3$ with $|\sigma|\leq 1$
and $x\in \R^3$ the following estimates
hold:
\begin{eqnarray*}
|D^\sigma (\nu (x)-\nu_\infty)| 
&\leq& 
C (1+|x|)^{-(1+|\sigma|)},\\
|D^\sigma (B-1) (x)| 
&\leq& 
C (1+|x|)^{-(2+|\sigma|)},\\
|D^\sigma \xi (x)|
&\leq& 
C (1+|x|)^{-(2+|\sigma|)}.
\end{eqnarray*}
In particular, the spacetime equipped with the metric
(\ref{metric_ax}) is asymptotically flat in the sense
that (\ref{bc_infinity_shift}) and, after a trivial change of coordinates,
also (\ref{bc_infinity}) holds.
\end{proposition}
{\bf Proof.} By definition of $G_1$,
$\lim_{|x|\to\infty} \nu(x)=\int\frac{M_1(y)}{|y|}\, dy$.
The first two estimates are established in
Lemma~\ref{Gwelldef}. As to the third one we observe that
by the boundary condition (\ref{bcxi_axis}) and Lemma~\ref{Gwelldef},
\[
|\xi(0,z)| \leq C |\partial_\rho h(0,z)| \leq \frac{C}{(1+|z|)^2}.
\]
By (\ref{rxi_eqn}) and the known
asymptotic behavior of the coefficients in that equation
which are given in terms of $\nu$ and $h$ and their derivatives,
\[
|\partial_\rho \xi (\rho,z)| \leq \frac{C}{1+\rho^3 + |z|^3},
\]
cf.\ Lemma~\ref{Gwelldef}.  Hence
\begin{eqnarray*}
|\xi(\rho,z)|
&\leq&
|\xi(0,z)| + \int_0^\rho |\partial_\rho \xi (s,z)|\, ds\\
&\leq&
\frac{C}{1+ |z|^2} + C \int_0^\infty \frac{ds}{1+s^3 + |z|^3} \leq \frac{C}{1+ |z|^2}
\end{eqnarray*}
which is the desired estimate for $\xi(\rho,z)$, provided $\rho < |z|$.
Since we already know that the metric under consideration
satisfies the full set of the Einstein equations
we can now use (\ref{xi_eqna}) and (\ref{xi_eqnb})
to see that also $\partial_z \xi$ is given in terms
of $\nu$ and $h$ and their derivatives and satisfies
the same decay estimate as $\partial_\rho \xi$. 
Starting from
\[
|\xi(\rho,z)| \leq
|\xi(\rho,\rho)| + \int_z^\rho |\partial_z \xi (\rho,s)|\, ds,
\]
we can use the decay of $\partial_z \xi$ to obtain the decay estimate
for $\xi(\rho,z)$
for $\rho \geq z \geq 0$ (or $\rho \geq -z \geq 0$),
and the proof is complete.
\prfe


\section{Regularity of axially symmetric functions}
\label{regularity}
\setcounter{equation}{0}


We call a function $f:\R^3 \to \R$ {\em axially symmetric}
if there exists a function $\tilde f:[0,\infty[ \times \R \to \R$
such that
\[
f(x) = \tilde f(\rho,z),\ \mbox{where}\
\rho=\sqrt{x_1^1+x_2^2}\ \mbox{and}\ z=x_3\ \mbox{for}\ x\in \R^3.
\]
In this section we collect some results on the relation
between the regularity properties of $f$ and those of $\tilde f$.
\begin{lemma} \label{axreg}
Let  $f:\R^3 \to \R$ be axially symmetric and
$f(x)=\tilde f(\rho,z)$ where 
$\tilde f:[0,\infty[ \times \R \to \R$. 
Let $k\in \{1,2,3\}$ and $\alpha\in]0,1[$. 
\begin{itemize}
\item[(a)]
$f\in C^k(\R^3)$ iff $\tilde f \in C^k([0,\infty[ \times \R)$
and all derivatives of $\tilde f$ of order up to $k$ which are
of odd order in $\rho$ vanish for $\rho=0$.
\item[(b)]
$f$ is H\"older continuous with exponent $\alpha\in]0,1[$
iff $\tilde f$ is.
\end{itemize}
\end{lemma}
{\bf Proof.}
As to part (a) let $f\in C^k(\R^3)$ be axially symmetric.
Then $f$ is even in $x_1$ and $x_2$ and 
$\tilde f(\rho,z)=f(\rho,0,z)$. This proves the ``only-if'' part.
For the ``if'' part one checks that the corresponding
derivatives of $f$, which exist for $\rho\neq 0$, extend
continuously to $\rho=0$.
As to part (b) one only needs to
observe that $x \mapsto \rho(x) = \sqrt{x_1^2 + x_2^2}$
is Lipschitz, since $|\nabla \rho (x)| = 1$.
\prfe

At several places in our analysis it is convenient
to extend functions of $(\rho,z)$ to negative values of
$\rho$.
\begin{lemma} \label{hoverrhoreg}
Let $h=h(\rho,z)\in C^4(\R^2)$ be odd in $\rho$ and
define 
\[
b(\rho,z):=\left\{
\begin{array}{ccl}
h(\rho,z)/\rho&,&\rho\neq 0,\\
\partial_\rho h(0,z)&,&\rho=0.
\end{array} \right.
\]
Then $b\in C^3(\R^2)$ and all derivatives of $b$ up to order $3$
which are of odd order in $\rho$ vanish for $\rho=0$.
By abuse of notation, $b\in C^3(\R^3)$.
\end{lemma}
{\bf Proof.}
The regularity of $b$ only needs to be checked at $\rho=0$.
Since $h$ is odd in $\rho$ it follows that $h(0,z)=\partial_{\rho\rho}h(0,z)=0$
for $z\in \R$. Hence as $\rho \to 0$,
\[
b(\rho,z) = \frac{1}{\rho}(h(\rho,z) - h(0,z)) \to \partial_\rho h(0,z),
\]
and by Taylor expansion,
\begin{eqnarray}
\partial_\rho b(\rho,z)
&=&
\frac{1}{\rho} \partial_\rho h(\rho,z) - \frac{1}{\rho^2} h(\rho,z)\nonumber \\
&=&
\frac{1}{\rho} \left(\partial_\rho h(0,z)+\partial_{\rho\rho} h(\tau,z)\rho\right)
\nonumber\\
&&
{}- \frac{1}{\rho^2} \left( h(0,z) + \partial_\rho h(0,z)\rho 
+\frac{1}{2} \partial_{\rho\rho} h(\sigma,z)\rho^2\right)\nonumber\\
&=&
\partial_{\rho\rho} h(\tau,z) - \frac{1}{2} \partial_{\rho\rho} h(\sigma,z)
\label{hoverrho}\\
&\to&
\frac{1}{2} \partial_{\rho\rho} h(0,z) =0 \nonumber
\end{eqnarray}
where $\sigma,\tau$ are between $0$ and $\rho$.
All other derivatives can be treated in a similar fashion,
where one should observe that
$\partial_z h(0,z)=0$. The regularity with respect to $x$
then follows by Lemma~\ref{axreg}. \prfe


\section{${\cal F}$ is well defined} \label{Fwelldef}
\setcounter{equation}{0}


As a first step we investigate the regularity properties
of the functions $\Phi_{jj},\ j=0,\ldots,3$, and of the
induced matter terms $M_1$, $M_2$.
\begin{lemma} \label{Phireg}
Let $\phi$ and $\psi$ satisfy the conditions ($\phi 1$) and
($\psi$) respectively.
\begin{itemize}
\item[(a)]
The functions $\Phi_{00}$ and $\Phi_{33}$ have derivatives
with respect to $\nu,\xi,\rho\in \R$ and $B \in ]1/2,3/2[$
up to order three and these are continuous in 
$\nu,\xi,B,\rho,\gamma,\lambda$. The same is true for 
$\Phi_{11}$ for derivatives up to order four.
\item[(b)]
For $(\nu,\xi,h;\gamma,\lambda)\in{\cal U}$,
$M_1 \in C^2(\R^3)$ and $M_2 \in C^{2,\alpha}(\R^2)$
are both compactly supported.
\end{itemize}
\end{lemma}
{\bf Proof.}
As to part (a) we note that differentiability
with respect to $\xi$ and $\rho$ is straight forward.
Concerning differentiability with respect to
$\nu$ and $B$ we observe that for $j=0,\ldots,3$ the
expression $\Phi_{jj}$
is differentiable once with respect
to the indicated variables, provided $\phi\in L^\infty_\mathrm{loc}$,
cf.\ the proof of \cite[Lemma~2.1]{Rein00}. Under the assumption
($\phi 1$) we can first differentiate twice before the change
to the integration variables $\eta$ and $s$ and obtain expressions
which are essentially of the same form as $\Phi_{jj}$, but
with $\phi'$ or $\phi''$ instead of $\phi$ so that the resulting expression
can be differentiated once more. The reason why $\Phi_{11}$
is one order more differentiable is that when differentiating
this expression with respect to $\nu$ or $B$ the integral with
respect to $s$ is preserved, its integrand is differentiated,
and the resulting expression is qualitatively of the same type
as $\Phi_{00}$ and can be differentiated three more times.

Part (b) follows since the functions $\nu, B, \xi$
which are now substituted into $\Phi_{jj}$ are all
at least in $C^{2,\alpha}(\R^3)$; the fact that $\xi$ is defined
only on the cylinder $Z_R$ does not matter here because the
integrals in the definitions of $\Phi_{jj}$ yield
functions with support in $Z_R$.
\prfe
We now show that ${\cal F}$ is well defined,
more precisely: 
\begin{lemma} \label{Gwelldef} 
Let $(\nu,\xi,h;\gamma,\lambda)\in{\cal U}$.
Then the following holds.
\begin{itemize}
\item[(a)]
$G_1=G_1(\nu,\xi,h;\gamma,\lambda)\in C^{3,\alpha}(\R^3)$ is axially
symmetric, even in $z=x_3$, and $\|(1+|x|) (G_1-a)\|_\infty$,
$\|(1+|x|)^2 \nabla G_1\|_\infty < \infty$, where $a=\int\frac{M_1(y)}{|y|}dy$.
\item[(b)]
$G_2=G_2(\nu,\xi,h;\gamma,\lambda)\in C^{4,\alpha}(\R^2)$ is odd in $\rho$,
even in $z$, and 
\[
\|(1+|(\rho,z)|) G_2\|_\infty,\, 
\|(1+|(\rho,z)|)^2 D^1 G_2\|_\infty,\, 
\|(1+|(\rho,z)|)^3 D^2 G_2\|_\infty < \infty.
\]
Moreover,
\[
\|(1+|(\rho,z)|)^2 (G_2/\rho)\|_\infty,\ 
\|(1+|(\rho,z)|)^3 D^1(G_2/\rho)\|_\infty < \infty.
\]
Here $D^j$ stands for any derivative of order $j$ with
respect to $(\rho,z)\in \R^2$.
\item[(c)]
$G_3=G_3(\nu,\xi,h;\gamma,\lambda)\in C^{2,\alpha}(Z_R)$ is axially
symmetric, even in $z=x_3$, and $\|G_3\|_{C^{2,\alpha}(Z_R)}< \infty$. 
\item[(d)]
${\cal F}(\nu,\xi,h;\gamma,\lambda)\in {\cal X}$.
\end{itemize}
\end{lemma}
{\bf Proof.}
As to part (a) the potential induced by the matter term $M_1$,
which is in $C^{1,\alpha}(\R^3)$ by Lemma~\ref{Phireg}~(b),
has the desired regularity and decay properties due to standard
regularity results in H\"older spaces, 
cf.~\cite[Thms.~10.2, 10.3]{liebloss}, and the decay of
$1/|x-y|$ and its derivatives together with the compact
support of $M_1$. As to the source term 
$g=\nabla(h/\rho)\cdot \nabla \nu$ of the second term in $G_1$
we notice that
$\nu\in {\cal X}_1$ and $h\in {\cal X}_2$ implies that
$g \in C^{1,\alpha}(\R^3)$ with $|g(x)|\leq C (1+|x|)^{-3-\beta}$,
in particular $g\in L^1 \cap L^\infty (\R^3)$. This implies
the regularity of the potential induces by $g$ and also its
decay:
\begin{eqnarray*}
\int\frac{|g(y)|}{|x-y|} dy
&\leq&
\int_{|x-y|\leq |x|/2} \ldots  + \int_{|x-y| > |x|/2}\\
&\leq&
C \int_{|x-y|\leq |x|/2} (1+|y|)^{-3-\beta}\frac{dy}{|x-y|} + 
\frac{2}{|x|} \int |g(y)|\, dy\\
&\leq&
C (1+|x|/2)^{-3-\beta}\int_{|x-y|\leq |x|/2} \frac{dy}{|x-y|} + \frac{C}{|x|}
\leq
\frac{C}{|x|}
\end{eqnarray*}
for large $|x|$ as desired; for the gradient of the potential induced by $g$
we argue completely analogously.

As to part (b) we first recall that $M_2=M_2(\rho,z)$ is even in
$\rho$, and the actual source term $\rho M_2$ is odd, compactly
supported, and by Lemma~\ref{Phireg}~(b) and 
Lemma~\ref{axreg}~(b), $M_2\in C^{2,\alpha}(\R^2)$.
Hence $G_2 \in C^{4,\alpha}(\R^2)$ is odd in $\rho\in \R$.
As to the decay of $G_2$ let $\mathrm{supp}\, M_2 \subset B_R(0)\subset\R^2$.
Then for $|(\rho,z)| \geq 2 R$ and 
$(\tilde \rho,\tilde z)\in\mathrm{supp}\, M_2$
the estimate 
\[
\left|\ln |(\rho-\tilde \rho,z-\tilde z)| -
\ln |(\rho,z)|\right| \leq \frac{2 R}{|(\rho,z)|}
\]
holds, and since $\int \tilde\rho M_2 = 0$ this implies that
\[
|G_2(\rho,z)| = \left|G_2(\rho,z)- 
4 \int\ln |(\rho,z)|\, \tilde\rho M_2(\tilde \rho,\tilde z)\,
d\tilde z\,d\tilde\rho\right|
\leq \frac{C}{|(\rho,z)|};
\]
the estimates for the derivatives of $G_2$ follow along
the same lines. Finally, 
$\partial_\rho(G_2/\rho)=-G_2/\rho^2 + \partial_\rho G_2 /\rho$
which implies that
\[
|\partial_\rho(G_2/\rho)(\rho,z)| \leq 
\frac{C}{|(\rho,z)| \rho^2} + \frac{C}{|(\rho,z)|^2 |\rho|}.
\]
This yields the asserted decay when $|\rho|$ becomes large.
But we can also use (\ref{hoverrho}) to see that
$|\partial_\rho(G_2/\rho)(\rho,z)|\leq C/|z|^3$.
Both estimates together yield the asserted decay for  
$\partial_\rho(G_2/\rho)$, and the decay
for $G_2/\rho$ and $\partial_z(G_2/\rho)$
can be dealt with similarly.

In order to prove part (c) we observe that
(\ref{gdef}) and the regularity
of $\nu$ and $h$ imply that $g$ and hence
$G_3\in C^{2,\alpha}(Z_R)$. By construction,
$\partial_\rho G_3 = g$.
Since $h$ is odd in $\rho$ we find that
\[
h(0,z)=\partial_z h(0,z)=\partial_{zz} h(0,z)=\partial_{\rho\rho} h(0,z)=0,
\]
which implies that $g(0,z)=0$. Thus by Lemma~\ref{axreg},
$G_3\in C^{2,\alpha}(Z_R)$, and the proof is complete.
\prfe


\section{${\cal F}$ is continuous and continuously differentiable with respect to
$\nu,h,\xi$}
\label{Fdiff}
\setcounter{equation}{0}


In this section we give some details of the proof
of the following result:

\begin{lemma} \label{frechet}
The mappings
\[
G_i : {\cal U} \to {\cal X}_i,\ i=1,2,3
\]
are continuous and continuously Fr\'{e}chet differentiable
with respect to $\nu$, $h$, and $\xi$.
\end{lemma}
{\bf Proof.} We only show the differentiability assertion
and focus on $G_1$. Defining 
$\Phi = \Phi_{00}+\gamma \Phi_{11} +\gamma \Phi_{33}$
we consider the differentiability only with respect to $\nu$,
and neglecting the dependence on the remaining variables
we look at the prototype mapping
\[
G : {\cal V} \to {\cal X}_1,\ 
G(\nu)(x) := \int_{\R^3}\frac{\Phi(\nu(y))}{|x-y|}dy,
\]
where ${\cal V}\subset {\cal X}_1$ is open,
$\Phi\in C^3 (\R)$ and $\Phi\circ\nu$ has support in a
fixed ball for all $\nu\in{\cal V}$. Our first claim is that
$G$ has the Fr\'{e}chet derivative
\[
[DG(\nu)\delta \nu](x) = \int_{\R^3}\frac{\Phi'(\nu(y))\delta\nu(y)}{|x-y|}dy,\
\nu \in {\cal V},\ \delta\nu\in {\cal X}_1.
\]
In order to prove this claim we need to show that 
for $\nu \in {\cal V}$ there exists $\epsilon >0$ such that
for $\delta\nu\in B_\epsilon(0)\subset {\cal X}_1$,
\[
||G(\nu+\delta\nu)-G(\nu) - DG(\nu)\delta\nu||_{{\cal X}_1} = 
\mathrm{o}(||\delta\nu||_{{\cal X}_1}).
\]
The support property and the standard elliptic estimate imply that
\begin{eqnarray*}
&&
||G(\nu+\delta\nu)-G(\nu) - DG(\nu)\delta\nu||_{{\cal X}_1} \\
&&
\qquad \qquad \leq 
C\, ||G(\nu+\delta\nu)-G(\nu) - DG(\nu)\delta\nu||_{C^{3,\alpha}(\R^3)}\\
&&
\qquad \qquad \leq
C\, ||\Phi(\nu+\delta\nu)-\Phi(\nu) - \Phi'(\nu)\delta\nu||_{C^{1,\alpha}(\R^3)}\\
&&
\qquad \qquad \leq
C\, ||\Phi(\nu+\delta\nu)-\Phi(\nu) - \Phi'(\nu)\delta\nu||_{C^{2}_b(\R^3)}.
\end{eqnarray*}
Clearly,
\[
||\Phi(\nu+\delta\nu)-\Phi(\nu) - \Phi'(\nu)\delta\nu||_\infty
=\mathrm{o}(||\delta\nu||_\infty) \leq \mathrm{o}(||\delta\nu||_{{\cal X}_1}).
\]
We need to establish analogous estimates for expressions where
we take derivatives with respect to $x$ up to second order of
the left hand side. Let $i,j\in \{1,2,3\}$. Then
\begin{eqnarray*}
\partial_{x_i}\left(\Phi(\nu+\delta\nu)-\Phi(\nu) - \Phi'(\nu)\delta\nu\right)
&=&
\left(\Phi'(\nu+\delta\nu) -\Phi'(\nu)\right)\, \partial_{x_i}\delta\nu\\
&&
+\left(\Phi'(\nu+\delta\nu) -\Phi'(\nu) - \Phi''(\nu)\delta\nu\right)
\,\partial_{x_i} \nu
\end{eqnarray*}
where both terms on the right are $\mathrm{o}(||\delta\nu||_{{\cal X}_1})$.
Similarly,
\begin{eqnarray*}
&&
\partial_{x_i x_j}\left(\Phi(\nu+\delta\nu)-\Phi(\nu) - \Phi'(\nu)\delta\nu\right)\\
&&
\qquad\qquad\qquad
=
\left(\Phi''(\nu+\delta\nu) -\Phi''(\nu) - \Phi'''(\nu)\delta\nu\right)
\,\partial_{x_i} \nu\,\partial_{x_j} \nu\\
&&
\qquad\qquad\qquad\quad
+ 
\left(\Phi''(\nu+\delta\nu) -\Phi''(\nu)\right)\, 
\left(\partial_{x_i} \nu\,\partial_{x_j}\delta \nu + 
\partial_{x_j} \nu\,\partial_{x_i}\delta \nu\right)\\
&&
\qquad\qquad\qquad\quad
+ 
\left(\Phi'(\nu+\delta\nu) -\Phi'(\nu) - \Phi''(\nu)\delta\nu\right)
\,\partial_{x_i x_j} \nu\\
&&
\qquad\qquad\qquad\quad
+ 
\left(\Phi'(\nu+\delta\nu) -\Phi'(\nu)\right)\,\partial_{x_i x_j} \delta\nu
+ \Phi''(\nu+\delta\nu)\,\partial_{x_i}\delta\nu\,\partial_{x_j}\delta \nu,
\end{eqnarray*}
and all the terms appearing on the right are
$\mathrm{o}(||\delta\nu||_{{\cal X}_1})$. This proves the differentiability
assertion for $G$. As to the continuity of this derivative,
\begin{eqnarray*}
||DG(\nu)- DG(\tilde\nu)||_{L({\cal X}_1,{\cal X}_1)}
&=&
\sup_{||\delta\nu||_{{\cal X}_1}\leq 1}
\left\|\int_{\R^3}
\frac{\left[(\Phi'(\nu)-\Phi'(\tilde\nu))\,\delta\nu\right](y)}
{|\cdot -y|} dy\right\|_{{\cal X}_1}\\
&\leq& 
C \,\sup_{||\delta\nu||_{{\cal X}_1}\leq 1}
||(\Phi'(\nu)-\Phi'(\tilde\nu))\,\delta\nu||_{C^{1,\alpha}(\R^3)}\\
&\leq& 
C \,||\Phi'(\nu)-\Phi'(\tilde\nu)||_{C^2_b(\R^3)} \to 0
\ \mbox{as}\ \tilde \nu \ \to \nu \ \mbox{in}\ {\cal X}_1.
\end{eqnarray*}
These arguments prove the continuous Fr\'{e}chet differentiability
of the first part of $G_1$ with respect to $\nu$. The derivatives
with respect to $h$ or $\xi$ can be dealt with in exactly the same manner.
The source term in the potential which represents the second part
of $G_1$ can be expanded explicitly in powers of $\delta h$ and $\delta\nu$
which together with the standard elliptic estimate proves the assertion
for that term; note that both $B$ and $B+\delta h/\rho$
are bounded away from $0$. 

The mapping $G_2$ is treated in the same way as our
prototype $G$ above, except that we have to estimate the source term
including its third order derivatives, observing that
$\Phi_{11}$ has derivatives up to order four with respect to 
$\nu, B, \xi$.

The mapping $G_3$ is easier since the term $g$ defined in
(\ref{gdef}) can be expanded explicitly in powers of $\delta \nu$
and $\delta h$ where again we observe that the denominator
in that expression is bounded away from $0$. \prfe


\section{$D{\cal F}(U_N,0,0;0,0)$ is one-to-one and onto}
\label{onetooneonto}
\setcounter{equation}{0}

We recall from Section~\ref{sec_res} and Eqn.~(\ref{kernel}) that in 
order to prove that the map $L$ is one-to-one it remains to show that 
$g=0$ is the only solution of 
\begin{equation}\label{kernel2}
\Delta g = 4 \pi a_N g,\ g(0)=0,
\end{equation}
in the space ${\cal X}_1$.
Inspired by the method in \cite{Rein00} we expand $g$ into spherical harmonics 
$Y_{lm},\,l\in\N_0,\,m=-l,...,l,$ where we use the notation of \cite{J};
for a more mathematical reference on spherical harmonics see \cite{mueller}. 
Denote by $(r,\theta,\varphi)$ and $(s,\tau,\psi)$ the spherical 
coordinates of a point $x\in\R^3$ and $y\in\R^3$ respectively. 
For $l\in\N_0$ and $m=-l,\ldots,l$ we define 
\begin{equation}\label{glm}
g_{lm}(r):=\frac{1}{r^2}\int_{|x|=r}Y_{lm}^{*}(\theta,\varphi)\,g(x)\,dS_x.
\end{equation}
The symmetry assumptions in the function space ${\cal X}_1$ imply that
$g_{1-1}=g_{10}=g_{11}=0$, since up to multiplicative constants
the spherical harmonics with $l=1$ are given by $\sin\theta e^{\pm i \varphi}$
and $\cos \theta$. To proceed,
we use the following expansion, cf.\ \cite{J},
\[
\frac{1}{|x-y|}=
\sum_{l=0}^{\infty}\sum_{m=-l}^{l}\frac{4\pi}{2l+1}\frac{r_{<}^{l}}
{r_{>}^{l+1}}Y_{lm}^{*}(\tau,\psi)\,Y_{lm}(\theta,\varphi),
\]
where $r_{<}:=\min{(r,s)}$ and $r_{>}:=\max{(r,s)}$. 
In view of (\ref{kernel2}),
\begin{eqnarray*}
g_{lm}(r)
&=&
- \frac{1}{r^2}\int_{\R^3}\int_{|x|=r}\frac{1}{|x-y|}
Y_{lm}^{*}(\theta,\varphi)\,dS_{x} a_{N}(s)\,g(y)dy \\
&=&
- \frac{4\pi}{2l+1}\int_{0}^{\infty} a_{N}(s)\frac{r_{<}^{l}}{r_{>}^{l+1}}
\int_{|y|=s}Y_{lm}^{*}(\tau,\psi)\,g(y)\,dS_{y}ds \\
&=&
-\frac{4\pi}{2l+1}\int_{0}^{\infty} a_{N}(s)\frac{r_{<}^{l}}{r_{>}^{l+1}}
s^2g_{lm}(s)ds \\
&=&
-\frac{4\pi}{2l+1}\left(\int_{0}^{r} a_{N}(s)\frac{s^{l+2}}{r^{l+1}}g_{lm}(s)\,ds
+\int_{r}^{\infty} a_{N}(s)\frac{r^{l}}{s^{l-1}}g_{lm}(s)\,ds\right).
\end{eqnarray*}
By a straightforward computation we find that $g_{lm}$ satisfies the equation
\begin{equation} \label{glmeqn}
\left(r^2 g_{lm}'\right)' =\left(l(l+1)+4\pi\, r^2 a_{N}(r)\right)\,g_{lm},
\end{equation}
where prime denotes a derivative with respect to $r$.

We use this to show that $g_{00}=0$ as follows.
We define $w(r):=\sup_{0\leq s\leq r} |g_{00}'(s)|$ 
so that $|g_{00}(r)| \leq r w(r)$; at this point it becomes essential
that $g(0)=g_{00}(0)=0$.
Now (\ref{glmeqn}) can be integrated to yield the Gronwall estimate
\[
w(r) \leq 4 \pi \int_0^r s |a_N (s)|\, w(s)\, ds,\ r\geq 0,
\]
so that $w=0$ and hence $g_{00}=0$ as desired.

It therefore remains to consider $g_{lm}$ with $l\geq 2$. For these we 
prove the following auxiliary result.   

\begin{lemma}\label{lemmau} 
Let $a\in C_c([0, \infty[)$
and $\lambda >0$ be such that $\lambda+4\pi r^2 a(r)>0$ for
$r\in [0, \infty[$.
Let $u\in C^2([0, \infty[)$
be a bounded solution to
\begin{equation}\label{ueq}
(r^2 u')' = (\lambda+4\pi r^2 a(r))u.
\end{equation}
Then $u=0$.
\end{lemma}
{\bf Proof.} 
We fix $r_a>0$ such that $a(r)=0$ for $r\ge r_a$. Multiplying (\ref{ueq}) with $u$
and
integrating by parts we obtain for $r>0$,
\begin{eqnarray}
\int_0^r (\lambda +4\pi s^2 a(s))\,u^2(s)\,ds 
&=& 
\int_0^r (s^2 u'(s))'\, u(s)\, ds\nonumber\\ 
& = & 
r^2 u'(r)\,u(r)-\int_0^r s^2 (u'(s))^2\,ds. \label{ibprel}
\end{eqnarray}
Now if there exists $r_0>0$ so that $u(r_0)=0$ or $u'(r_0)=0$
then (\ref{ibprel}) implies that $u(r)=u'(r)=0$ for $r\in [0, r_0]$.
The unique solvability of (\ref{ueq}) for $r\ge r_0$ then shows that $u=0$
as claimed.

So we assume now that $u(r)\neq 0$ and $u'(r)\neq 0$ for $r>0$.
Since (\ref{ueq}) is invariant under $u\to -u$, we may suppose that
$u(r)>0$ and $u'(r)>0$ for all $r\in ]0, \infty[$;
note that (\ref{ibprel}) enforces $uu'>0$ on $]0, \infty[$. For $r\ge r_a>0$
(\ref{ueq}) simplifies to $(r^2 u')'=\lambda u$, which has the solution
\[ 
u(r)=\frac{(l+1)\, u(r_a)+r_a u'(r_a)}{(2l+1)}\bigg(\frac{r}{r_a}\bigg)^l
   +\frac{l\, u(r_a)-r_a u'(r_a)}{(2l+1)}\bigg(\frac{r_a}{r}\bigg)^{l+1}. 
\]
Therefore
$u$ is unbounded which is a contradiction.
\prfe

Since $g\in {\cal X}_1$, Eqn.~(\ref{glm}) implies that $g_{lm}$ 
is bounded. Due to ($\phi 3$) we can apply
Lemma \ref{lemmau} to conclude that
$g_{lm}= 0$ for all $l\geq 2$, and thus $g=0$ as desired. 

We now prove the compactness result which was needed
to show that $L$ is onto.
\begin{lemma} \label{compactness}
The mapping $K: {\cal X}_1\to {\cal X}_1$,
\[ 
(Kw)(x)=\int_{\R^3}\frac{a_N(y)\,w(y)}{|x-y|} dy 
\]
is compact.
\end{lemma}
We remark that the operator $L_1$ has the form
$L_1(\delta\nu)(x)=-K(\delta\nu)(x)+K(\delta\nu)(0)$
and is compact if $K$ is, since the mapping $\nu \mapsto \nu(0)$
is continuous on ${\cal X}_1$.

\noindent
{\bf Proof.} 
First we observe that the mapping
\[
u \mapsto \int_{\R^3}\phi'\left(\frac{1}{2}|v|^2 + u\right)\, dv
= 2 \sqrt{2} \pi \int_u^\infty \phi'(E)\, \sqrt{E-u}\, dE
\]
is in $C^2 (\R)$, and since $U_N\in C^2(\R^3)$
the function $a_N$ is in
$C^2_c (\R^3)$. Hence $a_N w \in C^{1,1/2}(\R^3)$
for any $w\in {\cal X}_1$, and since $\alpha < 1/2$ the mapping
$K$ is well defined.

We fix a function $\chi\in C_c^\infty(\R^3)$ such that $0\le\chi\le 1$,
$\chi(x)=1$ for $|x|\le 1$, and $\chi(x)=0$ for $|x|\ge 2$. 
Let $\chi_R(x)=\chi(x/R)$
for $R>0$ and define
\[ 
(K_R w)(x)=\chi_R(x)(Kw)(x). 
\]
We show that $K_R\to K$ in the operator norm as $R\to\infty$. 
To this end, let $\zeta_R = 1-\chi_R$ so that for
$w\in {\cal X}_1$ and $x\in \R^3$,
\begin{equation}\label{kapprox}
(K w - K_R w)(x) = \zeta_R(x) (Kw)(x),
\end{equation}
and the latter vanishes for $|x|\leq R$. 
Now let $\|w\|_{{\cal X}_1}\leq 1$. For $\sigma\in \N_0^3$ with
$|\sigma| \leq 3$ it follows that
\begin{eqnarray*}
\left|D^\sigma (K w - K_R w)(x)\right|
&\leq& 
\zeta_R(x) \left|D^\sigma (Kw)(x)\right| \\
&&
{}+ \sum_{0<\tau \leq \sigma}
\left|c_{\tau}D^{\tau}\zeta_R(x)D^{\sigma-\tau} (Kw)(x)\right|\\
&\leq&
{\bf 1}_{\{|x|\ge R\}}\frac{C}{|x|} + 
C \sum_{0<\tau \leq \sigma}\frac{1}{R} \left|D^{\sigma-\tau}(Kw)(x)\right|
\leq \frac{C}{R};
\end{eqnarray*}
constants denoted by $C$ do not depend on $x$ or $R$.
In order to estimate the H\"older norm of $D^\sigma (K w - K_R w)$
for $|\sigma|=3$ we take $x,\tilde x \in\R^3$ with $|\tilde x|\geq|x|$
and again apply the product rule to the expression (\ref{kapprox}).
Adding and subtracting terms we have to estimate expressions like
\[
\left|(D^\tau \zeta_R(x)- D^\tau \zeta_R(\tilde x)) D^{\sigma-\tau}(Kw)(x)\right|
\leq \frac{C}{R}|x-\tilde x|
\]
and terms like the following:
\[
\left|D^\tau \zeta_R(x)\right|\int_{\R^3}\left|D^{\sigma-\tau}|y|^{-1}\right|
\left|(a_N w)(x-y)-(a_N w)(\tilde x-y)\right|\, dy;
\]
if $|\sigma-\tau|=3$ we throw one derivative onto $a_N w$.
The latter quantity together with its first order derivatives
is H\"older continuous. The factor in front of the integral
vanishes for $|x|\leq R$, so we need only consider
$|\tilde x| \geq |x| \geq R$. Since the domain of integration
extends only over $y$ with $|y-x|\leq R_N$ or  
$|y-\tilde x|\leq R_N$ we can on the domain of integration 
estimate $|y|\geq |x|-|x-y| \geq R - R_N \geq R/2$
or analogously with $\tilde x$ instead of $x$, where we assume
that $R>2 R_N$. Since $\left|D^{\sigma-\tau}|y|^{-1}\right|\leq |y|^{-j}$
with $j\geq 1$ the term under consideration can be estimated
by $C R^{-1} |x-\tilde x|^\alpha$ and altogether we conclude
that
\[
\|K w - K_R w\|_{C^{3,\alpha}(\R^3)} \leq C/R.
\]
Recalling the definition of the norm
$\|\cdot\|_{{\cal X}_1}$ we see that 
the following chain of estimates finally shows
that $K_R \to K$
in the corresponding operator norm as desired:
\begin{eqnarray*}
 |\nabla (Kw)(x)-\nabla (K_R w)(x)|
&\le&
\zeta_R(x)\int_{\R^3}\frac{|a_N(y) w(y)|}{|x-y|^2}dy\\
&&
{}+ R^{-1}|\nabla\chi(x/R)|
\int_{\R^3}\frac{|a_N(y) w(y)|}{|x-y|}dy \\ 
&\le& 
{\bf 1}_{\{|x|\ge R\}}\,\frac{C}{|x|^2}
+R^{-1}\,{\bf 1}_{\{R\le |x|\le 2R\}} \frac{C}{|x|} \\
&\le& 
C (1+|x|)^{-\beta-1}R^{-(1-\beta)}.
\end{eqnarray*}
To complete the proof we have to show that $K_R$ is compact for
any $R>0$ on the space ${\cal X}_1$. First the fact that
$a_N \in C^2_c(\R^3)$ implies that
\[
K_R :  C^{3,\alpha}(\R^3) \to C^{3,1/2}(\R^3)
\]
is continuous, and the same is true for
\[
K_R :  C^{3,\alpha}(\R^3) \to C^{3,1/2}(\overline{B}_{3 R}(0))
\]
where we note that all the functions $K_R w$ with
$w\in C^{3,\alpha}(\R^3)$ are supported in $B_{3 R}(0)$.
Since $\alpha < 1/2$ the embedding
\[
C^{3,1/2}(\overline{B}_{3 R}(0))\hookrightarrow
C^{3,\alpha}(\overline{B}_{3 R}(0))
\]
is compact, and because of the support property we conclude that
\[
K_R : {\cal X}_1 \to {\cal X}_1
\]
is compact; on $\nabla K_R w$ the weight $(1+|x|)^{1+\beta}$
only amounts to multiplication with a bounded function.
\prfe


\section{Discussion of Condition ($\phi 3$)}
\label{an_cond}
\setcounter{equation}{0}

In this section we investigate Condition ($\phi 3$)
for the case of the polytropic steady states (\ref{poly}).
We first allow for the general range $k\in ]-1/2, 7/2[$
of polytropic exponent.
Using the elementary integration formula
\begin{equation}\label{elint}
\int_{\R^3} \Big(s-\frac{1}{2}|v|^2\Big)_+^k\,dv
=(2\pi)^{3/2}\frac{\Gamma(k+1)}{\Gamma(k+\frac{5}{2})}\,s_+^{k+\frac{3}{2}},
\quad s\in\R,
\end{equation}
the Poisson equation in ($\phi 2$) is found to be
\[
\frac{1}{r^2}{(r^2 U'_N)}'=
4\pi(2\pi)^{3/2}\frac{\Gamma(k+1)}{\Gamma(k+\frac{5}{2})}
   \,(E_0-U_N)_+^{k+\frac{3}{2}} 
\]
for $U_N=U_N(r)$. According to \cite{RR00} there exists a solution 
$U_N$ such that $U_N(0)<E_0$, $U'_N(0)=0$, $U_N(R_N)=E_0$, 
$U_N(r) > E_0$ for $r>R_N$,
and $U'_N(r)>0$ for $r\in ]0, R_N[$. For $z:=E_0-U_N$ this means that
\[ 
-\frac{1}{r^2}{(r^2 z')}'=4\pi c_n\,z_+^n,\quad 
\mbox{where}\ n:=k+\frac{3}{2}\in ]1, 5[,
\quad c_n:=(2\pi)^{3/2}\frac{\Gamma(k+1)}{\Gamma(k+\frac{5}{2})}, 
\]
and furthermore $z(0)>0$, $z'(0)=0$, $z(R_N)=0$,
and $z'(r)<0$ for $r\in ]0, R_N[$. In terms of $z$
the function $a_N$ from ($\phi 3$) reads
\[ 
a_N(r)=-(2\pi)^{3/2}\frac{k\Gamma(k)}{\Gamma(k+\frac{3}{2})}
   \,z(r)_+^{k+\frac{1}{2}}=-n\,c_n\,z(r)_+^{n-1}, 
\]
where once more (\ref{elint}) was used. Thus condition ($\phi 3$) 
is equivalent to
\begin{equation}\label{stabcon2}
   4\pi n\,c_n\,r^2 z(r)_+^{n-1}<6.
\end{equation}
Now consider the function $\zeta(s):=z(\alpha s)$ for $\alpha:=(4\pi c_n)^{-1/2}$.
It is found to satisfy the Emden-Fowler equation
\begin{equation}\label{phiODE}
 -\frac{1}{s^2}{(s^2 \zeta')}'=\zeta_+^n
\end{equation}
and $\zeta(0)>0$, $\zeta'(0)=0$, $\zeta(s_0)=0$ for $s_0:= R_N/\alpha$,
as well as $\zeta'(s)<0$ for $s\in ]0, s_0[$. In terms of $s=\alpha^{-1} r$
condition (\ref{stabcon2}) becomes
\begin{equation}\label{stabcon3}
   s^2\zeta(s)_+^{n-1}<\frac{6}{n}.
\end{equation}
The left-hand side can be conveniently expressed
by means of the dynamical systems representation
of (\ref{phiODE}). For, let
\[ 
U(t):=-\frac{s\zeta(s)^n}{\zeta'(s)}\ge 0, \quad 
V(t):=-\frac{s\zeta'(s)}{\zeta(s)}\ge 0, \quad 
t:=\ln s, 
\]
where we consider $t\in ]-\infty, \ln s_0[$. Then
\begin{equation}\label{UVsys}
   \dot{U}=U(3-U-nV),\quad\dot{V}=V(U+V-1),
\end{equation}
and $U(t)V(t)=s^2\zeta(s)_+^{n-1}$, which provides 
the relation to (\ref{stabcon3}).
Thus we have to verify that $U(t)V(t)<6/n$.
In the terminology of \cite[p.~501]{BPf}, where $m=0$, $\zeta$ is an $E$-solution
to (\ref{phiODE}). Thus \cite[Prop.~5.5]{BPf} implies that $(U(t), V(t))$
lies in the unstable manifold of the fixed point $P_3=(3, 0)$ of (\ref{UVsys}).
In particular, we have $\lim_{t\to -\infty} (U(t), V(t))=(3, 0)$.
Also note that $P_3$ is of saddle type with eigenvalues $-3$ and $2$;
the corresponding eigenvectors are $(1, 0)$ and $(-3n/5, 1)$.
Since the line $V=\frac{1}{n}(3-U)$ separates the regions $\dot{U}>0$
(below the line) and $\dot{U}<0$ (above the line),
a phase plane analysis reveals that we must always have $U(t)\le 3$,
so that $W(t):=U(t)V(t)\le 3V(t)$. In addition, it is calculated that 
$V$ and $W$ are solutions to the system
\begin{equation}\label{VW-system}
   \dot{V}=V(V-1)+W,\quad\dot{W}=W(2-(n-1)V),
\end{equation}
such that $\lim_{t\to -\infty} (V(t), W(t))=(0, 0)$. 
The origin is a fixed point of saddle type for (\ref{VW-system}), 
the eigenvalues are $-1$ and $2$
with corresponding eigenvectors $(1, 0)$ and $(1, 3)$. 
Note that $\dot{W}>0$ for $V<\frac{2}{n-1}$, $\dot{W}<0$
 for $V>\frac{2}{n-1}$, $\dot{V}>0$ above the curve
$V\mapsto V(1-V)$, and $\dot{V}<0$ below this curve. Since the curve
has unity slope at $V=0$, it follows that $(V(t), W(t))$,
lying in the unstable manifold of the origin, will be above the curve
for $t$ very negative. Then a phase plane analysis shows that this 
property persists for all times. In particular, we always have $\dot{V}>0$, 
and $W$ is increasing
until it reaches its maximal value for $t_0$ such that $V(t_0)=\frac{2}{n-1}$.
Thus our original problem of proving ($\phi 3$) is equivalent
to showing that $W(t_0)=\max W<6/n$. Thanks to the preceding observations
the parametrized curve $t\mapsto (V(t), W(t))$ for $t\in ]-\infty, t_0]$
can be rewritten as a curve $W=W(V)$ in the $(V, W)$-plane which solves
\begin{equation}\label{dWdV}
   \frac{dW}{dV}=\frac{W(2-(n-1)V)}{V(V-1)+W},
\end{equation}
and which is such that $W(0)=0$ and $W(\frac{2}{n-1})=\max W$.

\begin{lemma} If $k<7/2$ is sufficiently close to $7/2$,
then ($\phi 3$) holds for $\phi$ given by (\ref{poly}).
\end{lemma}
{\bf Proof.} If $W(V)<1<6/n$ for all $V\in ]0, \frac{2}{n-1}]$, 
then we are done. Hence we assume that $W(V_0)=1$ for some
$V_0\in ]0, \frac{2}{n-1}]$. 
Then $1=W(V_0)\le 3V_0$ yields $V_0\ge 1/3$.
Since $W(V)\ge 1$ for $V\ge V_0$, it follows that 
$V(V-1)+W=(V-1)^2+V+W-1\ge V$,
so that by (\ref{dWdV}),
\begin{eqnarray*}
\ln(\max W) 
& = & 
\int_{W=1}^{\max W}\frac{dW}{W}
\le\int_{V_0}^{\frac{2}{n-1}}\frac{(2-(n-1)\tilde{V})}{\tilde{V}}\,d\tilde{V}\\ 
& \le & 
\int_{1/3}^{\frac{2}{n-1}}\frac{(2-(n-1)\tilde{V})}{\tilde{V}}\,d\tilde{V}\\ 
& = & 
2\ln\Big(\frac{6}{n-1}\Big)-2+\frac{1}{3}(n-1).
\end{eqnarray*}
Therefore
\begin{equation}\label{hileq}
\max W\le\frac{36}{(n-1)^2}\exp\Big(\frac{1}{3}(n-1)-2\Big).
\end{equation}
At $n=5$ the relation
\[ 
\frac{9}{4}\,e^{-2/3}<\frac{6}{5} 
\]
holds. Hence it follows from (\ref{hileq}) that $\max W<6/n$
is verified for $n$ sufficiently close to $n=5$. \prfe

The method of proof for the preceding lemma can be refined as follows. 
Fix $A<6/n$. Then $W(V)<A$ for $V\in [0, \frac{2}{n-1}]$ would be acceptable. 
Hence we can assume that $W(V_0)=A$ for some $V_0\in ]0, \frac{2}{n-1}]$. 
Then $A=W(V_0)\le 3V_0$ shows that $V_0\ge A/3$. From $W(V)\ge A$ 
for $V\ge V_0$ we obtain
\begin{eqnarray*}
&&\ln(\max W)-\ln A \\
&&
\qquad\quad = 
\int_{W(V_0)}^{\max W}\frac{dW}{W}
\le\int_{V_0}^{\frac{2}{n-1}}\frac{(2-(n-1)\tilde{V})}
{\tilde{V}^2-\tilde{V}+A}\,d\tilde{V}\\ 
&&
\qquad\quad \le 
\int_{A/3}^{\frac{2}{n-1}}\frac{(2-(n-1)\tilde{V})}{\tilde{V}^2-\tilde{V}+A}
\,d\tilde{V} \\ 
&&
\qquad\quad =
\frac{5-n}{\sqrt{4A-1}}\bigg[\arctan\Big(\frac{5-n}{\sqrt{4A-1}(n-1)}\Big)
+\arctan\Big(\frac{3-2A}{3\sqrt{4A-1}}\Big)\bigg] \\ 
&& 
\qquad\qquad {}-\Big(\frac{n-1}{2}\Big)
\ln\Big(\frac{9[An^2-2(A+1)n+6+A]}{A(A+6)(n-1)^2}\Big).
\end{eqnarray*}
Therefore $\max W\leq \Phi_A(n)$ where
\begin{eqnarray*}
&&
\Phi_A(n) :=
A\, \Big(\frac{A(A+6)(n-1)^2}{9[An^2-2(A+1)n+6+A]}\Big)^{\frac{n-1}{2}}\\
&&
\quad \times \exp\Bigg(\frac{5-n}{\sqrt{4A-1}}
\bigg[\arctan\Big(\frac{5-n}{\sqrt{4A-1}(n-1)}\Big)
+\arctan\Big(\frac{3-2A}{3\sqrt{4A-1}}\Big)\bigg]\Bigg).
\end{eqnarray*}
For different $A$ it can be checked (e.g.~using Maple) for which values
$n\in ]1, \min\{6/A, 5\}[$ the relation $\Phi_A(n)<6/n$ is verified.
Taking $A=1$ we get at least $n\in [2.6, 5[$,
for $A=6/5$ we get at least $n\in [2.35, 4.85]$,
and for $A=2$ we get at least $n\in [2.1, 2.5]$.
In summary, the desired relation $\max W<6/n$ can be obtained
for at least $n\in [2.1, 5[$, which corresponds to at least
$k\in [0.6, 3.5[$ in (\ref{poly}). Notice however that the regularity
assumption on $\phi$ requires $k>2$.


\section{The field equations hold}
\label{allfield}
\setcounter{equation}{0}


For a metric of the form (\ref{metric_ax}) the components
$00$, $11$, $12$, $22$, and $33$ of the field equations
are nontrivial. We have so far obtained a solution
$\nu,B,\xi$ of the reduced system (\ref{nu_eqn}), (\ref{B_eqn}),
(\ref{rxi_eqn}) where the appearing components of the energy
momentum tensor are induced by a phase space density $f$
which satisfies the Vlasov equation (\ref{vlasov_gen}).
We define $E_{\alpha\beta}:= G_{\alpha\beta} - 8 \pi c^{-4} T_{\alpha\beta}$
so that the Einstein field equations become $E_{\alpha\beta}=0$. 
By (\ref{B_eqn}),
\begin{equation}\label{E11+E22}
E_{11} + E_{22} =0.
\end{equation} 
Using this information (\ref{nu_eqn})
says that
\[
\rho^2 B^2 E_{00} + c^2 e^{4\nu/c^2}E_{33} =0
\]
or
\begin{equation} \label{e00e33}
c^2 e^{4\nu/c^2} E^{00} + \rho^2 B^2 E^{33} =0.
\end{equation}
The Vlasov equation implies that 
$\nabla_\alpha T^{\alpha\beta} = 0$,
and $\nabla_\alpha G^{\alpha\beta} = 0$ due to the contracted Bianchi
identity where $\nabla_\alpha$ denotes the covariant
derivative corresponding to the metric
(\ref{metric_ax}). We want to use these relations
to show that the remaining components of $E_{\alpha\beta}$
vanish also, but there is a technical catch: The metric,
more specifically $\xi$, is only $C^2$. To overcome this complication
we approximate $\xi$ by $C^3$ functions $\xi_n$. The induced Einstein 
tensor $G_n^{\alpha \beta}$ again satisfies the Bianchi identity.
Taking $\beta=1$ and letting $n\to \infty$ we obtain the equation
\begin{equation} \label{e12z}
\partial_z E^{12}+
\left( 4 \partial_z\mu +\frac{\partial_z B}{B}\right)\, E^{12}
- \rho B e^{-2\xi} (B+\rho\partial_\rho B)\, E^{33}=0,
\end{equation} 
where (\ref{e00e33}) has been used to eliminate $E^{00}$
and we recall that $\xi=\nu/c^2 + \mu$.
Here $\partial_z E^{12}$ is at first a distributional derivative,
but since all other terms in the equation are continuous
this derivative indeed exists in the classical sense.
The same approximation maneuver can be performed for $\beta=2$
to obtain the equation
\begin{equation} \label{e12rho}
\partial_\rho E^{12}+
\left( 4 \partial_\rho\mu +\frac{1}{\rho}+\frac{\partial_\rho B}{B}\right)\, E^{12}
- \rho^2 B e^{-2\xi} \partial_z B\, E^{33}=0
\end{equation} 
which holds for $\rho>0$. However, if we multiply this equation with
$\rho$ we obtain an equation which holds for $\rho\geq 0$. This is because
$E^{12}(0,z)=0$ which is nothing but the boundary condition
(\ref{bcxi_axis}) on the axis of symmetry which we have incorporated into
our integration of (\ref{rxi_eqn}).
We eliminate $E^{33}$ from (\ref{e12z}), (\ref{e12rho}) and write
the resulting equation for $E^{12}$ in terms of
\[
X:=\rho e^{4\mu} B E^{12}.
\]
The result is the equation  
\[
\partial_\rho X - 
\frac{\rho \partial_z B}{B+\rho \partial_\rho B}\partial_z X = 0
\]
which again holds for $\rho\geq 0$.
Since $X(0,z)=0$ and since any characteristic curve of this
equation intersects the axis of symmetry $\rho=0$
we conclude that $X$ vanishes identically.
By (\ref{e12z}) the same is true for $E^{33}$ so that
$E^{12}=E^{33}=E^{00}=0$.
Finally we observe that by (\ref{rxi_eqn}),
\[
\left(1+\rho \frac{\partial_\rho B}{B}\right)\, \left(E_{11} - E_{22}\right)
+ \rho \frac{\partial_z B}{B} E_{12} = 0.
\]
Since $E_{12}=0$ this means that $E_{11}=E_{22}$, and with
(\ref{E11+E22}) we conclude that $E_{11}=E_{22}=0$,
and all the non-trivial field equations are satisfied.

\smallskip

\noindent
{\bf Acknowledgment.} The authors would like to thank
Marcus Ansorg for useful discussions.

\end{document}